# Automatic Partitioning of Database Applications

Alvin Cheung *
Samuel Madden
MIT CSAIL

{akcheung, madden}@csail.mit.edu

Owen Arden†
Andrew C. Myers
Department of Computer Science,
Cornell University

{owen, andru}@cs.cornell.edu

## ABSTRACT

Database-backed applications are nearly ubiquitous in our daily lives. Applications that make many small accesses to the database create two challenges for developers: increased latency and wasted resources from numerous network round trips. A well-known technique to improve transactional database application performance is to convert part of the application into stored procedures that are executed on the database server. Unfortunately, this conversion is often difficult. In this paper we describe Pyxis, a system that takes database-backed applications and automatically partitions their code into two pieces, one of which is executed on the application server and the other on the database server. Pyxis profiles the application and server loads, statically analyzes the code's dependencies, and produces a partitioning that minimizes the number of control transfers as well as the amount of data sent during each transfer. Our experiments using TPC-C and TPC-W show that Pyxis is able to generate partitions with up to $3\times$ reduction in latency and $1.7\times$ improvement in throughput when compared to a traditional non-partitioned implementation and has comparable performance to that of a custom stored procedure implementation.

## 1. INTRODUCTION

Transactional database applications are extremely latency sensitive for two reasons. First, in many transactional applications (e.g., database-backed websites), there is typically a hard response time limit of a few hundred milliseconds, including the time to execute application logic, retrieve query results, and generate HTML . Saving even a few tens of milliseconds of latency per transaction can be important in meeting these latency bounds. Second, longer-latency transactions hold locks longer, which can severely limit maximum system throughput in highly concurrent systems.

Stored procedures are a widely used technique for improving the latency of database applications. The idea behind stored procedures is to rewrite sequences of application logic that are interleaved with database commands (e.g., SQL queries) into parameterized blocks of code that are stored on the database server. The application then sends commands to the database server, typically on a separate physical machine, to invoke these blocks of code.

Stored procedures can significantly reduce transaction latency by avoiding round trips between the application and database servers. These round trips would otherwise be necessary in order to execute the application logic found between successive database commands. The resulting speedup can be substantial. For example, in a Java implementation of a TPC-C-like benchmark—which has relatively little application logic—running each TPC-C transaction as a stored procedure can offer up to a $3\times$ reduction in latency versus running each SQL command via a separate JDBC call. This reduction results in a $1.7\times$ increase in overall transaction throughput on this benchmark.

However, stored procedures have several disadvantages:

• **Portability and maintainability**: Stored procedures break a straight-line application into two distinct and logically disjoint code bases. These code bases are usually written in different languages and must be maintained separately. Stored-procedure languages are often database-vendor specific, making applications that use them less portable between databases. Programmers are less likely to be familiar with or comfortable in low-level—even arcane—stored procedure languages like PL/SQL or TransactSQL, and tools for debugging and testing stored procedures are less advanced than those for more widely used languages.

• **Conversion effort:** Identifying sections of application logic that are good candidates for conversion into stored procedures is tricky. In order to design effective stored procedures, programmers must identify sections of code that make multiple (or large) database accesses and can be parameterized by relatively small amounts of input. Weighing the relative merits of different designs requires programmers to model or measure how often a stored procedure is invoked and how much parameter data need to be transferred, both of which are nontrivial tasks.

• **Dynamic server load:** Running parts of the application as stored procedures is not always a good idea. If the database server is heavily loaded, pushing more computation into it by calling stored procedures will hurt rather than help performance. A database server's load tends to change over time, depending on the workload and the utilization of the applications it is supporting, so it is difficult for developers to predict the resources available on the servers hosting their applications. Even with accurate predictions they have no easy way to adapt their programs' use of stored procedures to a dynamically changing server load.

We propose that these disadvantages of manually generated stored procedures can be avoided by automatically identifying and extracting application code to be shipped to the database server. We implemented this new approach in Pyxis, a system that automatically partitions a database application into two pieces, one de-ployed on the application server and the other in the database server as stored procedures. The two programs communicate with each other via remote procedure calls (RPCs) to implement the semantics of the original application. In order to generate a partition, Pyxis first analyzes application source code using static analysis and then collects dynamic information such as runtime profile and machine loads. The collected profile data and results from the analysis are then used to formulate a linear program whose objective is to minimize, subject to a maximum CPU load, the overall latency due to network round trips between the application and database servers as well as the amount of data sent during each round trip. The solved linear program then yields a fine-grained, statement-level partitioning of the application's source code. The partitioned code is split into two halves and executed on the application and database servers using the Pyxis runtime.

---

*Supported by a NSF Fellowship
†Supported by a DoD NDSEG Fellowship

The main benefit of our approach is that the developer does not need to manually decide which part of her program should be executed where. Pyxis identifies good candidate code blocks for conversion to stored procedures and automatically produces the two distinct pieces of code from the single application codebase. When the application is modified, Pyxis can automatically regenerate and redeploy this code. By periodically re-profiling their application, developers can generate new partitions as load on the server or application code changes. Furthermore, the system can switch between partitions as necessary by monitoring the current server load.

Pyxis makes several contributions:

1. We present a formulation for automatically partitioning programs into stored procedures that minimize overall latency subject to CPU resource constraints. Our formulation leverages a combination of static and dynamic program analysis to construct a linear optimization problem whose solution is our desired partitioning.

2. We develop an execution model for partitioned applications where consistency of the distributed heap is maintained by automatically generating custom synchronization operations.

3. We implement a method for adapting to changes in real-time server load by dynamically switching between pre-generated partitions created using different resource constraints.

4. We evaluate our Pyxis implementation on two popular transaction processing benchmarks, TPC-C and TPC-W, and compare the performance of our partitions to the original program and versions using manually created stored procedures. Our results show Pyxis can automatically partition database programs to get the best of both worlds: when CPU resources are plentiful, Pyxis produces a partition with comparable performance to that of hand-coded stored procedures; when resources are limited, it produces a partition comparable to simple client-side queries.

The rest of the paper is organized as follows. We start with an architectural overview of Pyxis in Sec. 2. We describe how Pyxis programs execute and synchronize data in Section Sec. 3. We present the optimization problem and describe how solutions are obtained in Sec. 4. Sec. 5 explains the generation of partitioned programs, and Sec. 6 describes the Pyxis runtime system. Sec. 7 shows our experimental results, followed by related work and conclusions in Sec. 8 and Sec. 9.

## 2. OVERVIEW

Figure 1 shows the architecture of the Pyxis system. Pyxis starts with an application written in Java that uses JDBC to connect to the database and performs several analyses and transformations.[1] The analysis used in Pyxis is general and does not impose any restrictions on the programming style of the application. The final result is two separate programs, one that runs on the application server and one that runs on the database server. These two programs communicate with each other as necessary to implement the original application's semantics. Execution starts at the partitioned program on the application server but periodically switches to its counterpart on the database server, and vice versa. We refer to these switches as *control transfers*. Each statement in the original program is assigned a *placement* in the partitioned program on either the application server or the database server. Control transfers occur when a statement with one placement is followed by a statement with a different placement. Following a control transfer, the calling program blocks until the callee returns control. Hence, a single thread of control is maintained across the two servers.

---

[1] We chose Java due to its popularity in writing database applications. Our techniques can be applied to other languages as well.

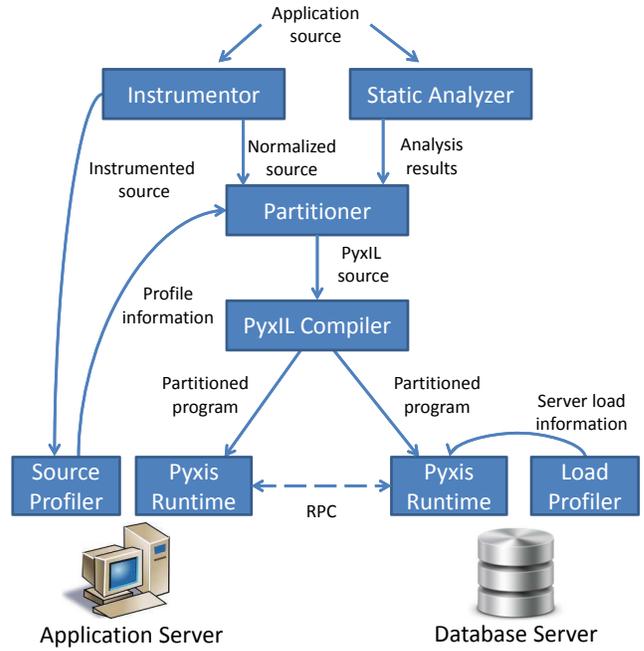

Figure 1: Pyxis architecture

Although the two partitioned programs execute in different address spaces, they share the same logical heap and execution stack. This program state is kept in sync by transferring heap and stack updates during each control transfer or by fetching updates on demand. The execution stack is maintained by the Pyxis runtime, but the program heap is kept in sync by explicit heap synchronization operations, generated using a conservative static program analysis.

**Static dependency analysis.** The goal of partitioning is to preserve the original program semantics while achieving good performance. This is done by reducing the number of control transfers and amount of data sent during transfers as much as possible. The first step is to perform an interprocedural static dependency analysis that determines the data and control dependencies between program statements. The data dependencies conservatively capture all data that may be necessary to send if a dependent statement is assigned to a different partition. The control dependencies capture the necessary sequencing of program statements, allowing the Pyxis code generator to find the best program points for control transfers.

The results of the dependency analysis are encoded in a graph form that we call a *partition graph*. It is a program dependence graph (PDG)[2] augmented with extra edges representing additional information. A PDG-like representation is appealing because it combines both data and control dependencies into a single representation. Unlike a PDG, a partition graph has a weight that models the cost of satisfying the edge's dependencies if the edge is partitioned so that its source and destination lie on different machines. The partition graph is novel; previous automatic partitioning approaches have partitioned control-flow graphs [34, 8] or dataflow graphs [23, 33]. Prior work in automatic parallelization [28] has also recognized the advantages of PDGs as a basis for program representation.

---

[2] Since it is interprocedural, it actually is closer to a *system dependence graph* [14, 17] than a PDG.

**Profile data collection.** Although the static dependency analysis defines the structure of dependencies in the program, the system needs to know how frequently each program statement is executed in order to determine the optimal partition; placing a "hot" code fragment on the database server may increase server load beyond its capacity. In order to get a more accurate picture of the runtime behavior of the program, statements in the partition graph are weighted by an estimated execution count. Additionally, each edge is weighted by an estimated latency cost that represents the communication overhead for data or control transfers. Both weights are captured by dynamic profiling of the application.

For applications that exhibit different operating modes, such as the browsing versus shopping mix in TPC-W, each mode could be profiled separately to generate partitions suitable for it. The Pyxis runtime includes a mechanism to dynamically switch between different partitionings based on current CPU load.

**Optimization as integer programming.** Using the results from the static analysis and dynamic profiling, the Pyxis partitioner formulates the placement of each program statement and each data field in the original program as an integer linear programming problem. These placements then drive the transformation of the input source into the intermediate language *PyxIL* (for PYXis Intermediate Language). The PyxIL program is very similar to the input Java program except that each statement is annotated with its placement, `:APP:` or `:DB:`, denoting execution on the application or database server. Thus, PyxIL compactly represents a distributed program in a single unified representation. PyxIL code also includes explicit heap synchronization operations, which are needed to ensure the consistency of the distributed heap.

In general, the partitioner generates several different partitionings of the program using multiple *server instruction budgets* that specify upper limits on how much computation may be executed at the database server. Generating multiple partitions with different resource constraints enables automatic adaptation to different levels of server load.

**Compilation from PyxIL to Java.** For each partitioning, the PyxIL compiler translates the PyxIL source code into two Java programs, one for each runtime. These programs are compiled using the standard Java compiler and linked with the Pyxis runtime. The database partition program is run in an unmodified JVM colocated with the database server, and the application partition is similarly run on the application server. While not exactly the same as running traditional stored procedures inside a DBMS, our approach is similar to other implementations of stored procedures that provide a foreign language interface such as PL/Java [1] and execute stored procedures in a JVM external to the DBMS. We find that running the program outside the DBMS does not significantly hurt performance as long as it is colocated. With more engineering effort, the database partition program could run in the same process as the DBMS.

**Executing partitioned programs.** The runtimes for the application server and database server communicate over TCP sockets us-

```
 1  class Order {
 2    int id;
 3    double[] realCosts;
 4    double totalCost;
 5    Order(int id) {
 6      this.id = id;
 7    }
 8    void placeOrder(int cid, double dct) {
 9      totalCost = 0;
10      computeTotalCost(dct);
11      updateAccount(cid, totalCost);
12    }
13    void computeTotalCost(double dct) {
14      int i = 0;
15      double[] costs = getCosts();
16      realCosts = new double[costs.length];
17      for (itemCost : costs) {
18        double realCost;
19        realCost = itemCost * dct;
20        totalCost += realCost;
21        realCosts[i++] = realCost;
22        insertNewLineItem(id, realCost);
23      }
24    }
25  }
```

**Figure 2: Running example**

ing a custom remote procedure call mechanism. The RPC interface includes operations for control transfer and state synchronization. The runtime also periodically measures the current CPU load on the database server to support dynamic, adaptive switching among different partitionings of the program. The runtime is described in more detail in Sec. 6.

## 3. RUNNING PYXIS PROGRAMS

Figure 2 shows a running example used to explain Pyxis throughout the paper. It is meant to resemble a fragment of the new-order transaction in TPC-C, modified to exhibit relevant features of Pyxis. The transaction retrieves the order that a customer has placed, computes its total cost, and deducts the total cost from the customer's account. It begins with a call to `placeOrder` on behalf of a given customer `cid` at a given discount `dct`. Then `computeTotalCost` extracts the costs of the items in the order using `getCosts`, and iterates through each of the costs to compute a total and record the discounted cost. Finally, control returns to `placeOrder`, which updates the customer's account. The two operations `insertNewLineItem` and `updateAccount` update the database's contents while `getCosts` retrieves data from the database. If there are $N$ items in the order, the example code incurs $N$ round trips to the database from the `insertNewLineItem` calls, and two more from `getCosts` and `updateAccount`.

There are multiple ways to partition the fields and statements of this program. An obvious partitioning is to assign all fields and statements to the application server. This would produce the same number of remote interactions as in the standard JDBC-based implementation. At the other extreme, a partitioning might place all statements on the database server, in effect creating a stored procedure for the entire `placeOrder` method. As in a traditional stored procedure, each time `placeOrder` is called the values `cid` and `dct` must be serialized and sent to the remote runtime. Other partitionings are possible. Placing only the loop in `computeTotalCost` on the database would save $N$ round trips if no additional communication were necessary to satisfy data dependencies. In general, assigning more code to the database server can reduce latency, but it also can put additional load on the database. Pyxis aims to choose partitionings that achieve the smallest latency possible using the current available resources on the server.

```
1   class Order {
2     :APP: int id;
3     :APP: double[] realCosts;
4     :DB: double totalCost;
5     Order(int id) {
6       :APP: this.id = id;
7       :APP: sendAPP(this);
8     }
9     void placeOrder(int cid, double dct) {
10      :APP: totalCost = 0;
11      :APP: sendDB(this);
12      :APP: computeTotalCost(dct);
13      :APP: updateAccount(cid, totalCost);
14    }
15    void computeTotalCost(double dct) {
16      int i; double[] costs;
17      :APP: costs = getCosts();
18      :APP: realCosts = new double[costs.length];
19      :APP: sendAPP(this);
20      :APP: sendNative(realCosts,costs);
21      :APP: i = 0;
22      for (:DB: itemCost : costs) {
23        double realCost;
24        :DB: realCost = itemCost * dct;
25        :DB: totalCost += realCost;
26        :DB: sendDB(this);
27        :DB: realCosts[i++] = realCost;
28        :DB: sendNative(realCosts);
29        :DB: insertNewLineItem(id, realCost);
30      }
31    }
32  }
```

**Figure 3: A PyxIL version of the Order class**

## 3.1 A PyxIL Partitioning

Figure 3 shows PyxIL code for one possible partitioning of our running example. PyxIL code makes explicit the placement of code and data, as well as the synchronization of updates to a distributed heap, but keeps the details of control and data transfers abstract. Field declarations and statements in PyxIL are annotated with a placement label (:APP: or :DB:). The placement of a statement indicates where the instruction is executed. For field declarations, the placement indicates where the authoritative value of the field resides. However, a copy of a field's value may be found on the remote server. The synchronization protocols using a conservative program analysis ensure that this copy is up to date if it might be used before the next control transfer. Thus, each object apparent at the source level is represented by two objects, one at each server. We refer to these as the APP and DB parts of the object.

In the example code, the field id is assigned to the application server, indicated by the :APP: placement label, but field totalCost is assigned to the database, indicated by :DB:. The array allocated on line 18 is placed on the application server. All statements are placed on the application server except for the for loop in computeTotalCost. When control flows between two statements with different placements, a *control transfer* occurs. For example, on line 21 in Fig. 3, execution is suspended at the application server and resumes at line 22 on the database server.

Arrays are handled differently from objects. The placement of an array is defined by its allocation site: that is, the placement of the statement allocating the array. This approach means the contents of an array may be assigned to either partition, but the elements are not split between them. Additionally, since the stack is shared by both partitions, method parameters and local variable declarations do not have placement labels in PyxIL.

## 3.2 State Synchronization

Although all data in PyxIL has an assigned placement, remote data may be transferred to or updated by any host. Each host maintains a local heap for fields and arrays placed at that host as well as a *remote cache* for remote data. When a field or array is accessed, the current value in the local heap or remote cache is used. Hosts synchronize their heaps using *eager batched updates*; modifications are aggregated and sent on each control transfer so that accesses made by the remote host are up to date. The host's local heap is kept up to date whenever that host is executing code. When a host executes statements that modify remotely partitioned data in its cache, those updates must be transferred on the next control transfer. For the local heap, however, updates only need to be sent to the remote host before they are accessed. If static analysis determines no such access occurs, no update message is required.

In some scenarios, eagerly sending updates may be suboptimal. If the amount of latency incurred by transferring unused updates exceeds the cost of an extra round trip communication, it may be better to request the data *lazily* as needed. In this work, we only generate PyxIL programs that send updates eagerly, but investigating hybrid update strategies is an interesting future direction.

PyxIL programs maintain the above heap invariants using explicit synchronization operations. Recall that classes are partitioned into two partial classes, one for APP and one for DB. The sendAPP operation sends the APP part of its argument to the remote host. In Fig. 3, line 7 sends the contents of id while line 26 sends totalCost and the array reference realCosts. Since arrays are placed dynamically based on their allocation site, a reference to an array may alias both locally and remotely partitioned arrays. On line 20, the contents of the array realCosts allocated at line 18 are sent to the database server using the sendNative operation[3]. The sendNative operation is also used to transfer unpartitioned native Java objects that are serializable. Like arrays, native Java objects are assigned locations based on their allocation site.

Send operations are batched together and executed at the next control transfer. Even though the sendDB operation on line 26 and the sendNative operation on line 28 are inside a for loop, the updates will only be sent when control is transferred back to the application server.

## 4. PARTITIONING PYXIS CODE

Pyxis finds a partitioning for a user program by generating partitions with respect to a representative workload that a user wishes to optimize. The program is profiled using this workload, generating inputs that partitioning is based upon.

### 4.1 Profiling

Pyxis profiles the application in order to be able to estimate the size of data transfers and the number of control transfers for any particular partitioning. To this end, statements are instrumented to collect the number of times they are executed, and assignment expressions are instrumented to measure the average size of the assigned objects. The application is then executed for a period of time to collect data. Data collected from profiling is used to set weights in the partition graph.

This profile need not perfectly characterize the future interactions between the database and the application, but a grossly inaccurate profile could lead to suboptimal performance. For example, with inaccurate profiler data, Pyxis could choose to partition the program where it expects few control transfers, when in reality the program might exercise that piece of code very frequently. For this reason, developers may need to re-profile their applications if the workload changes dramatically and have Pyxis dynamically switch among the different partitions.

---

[3]For presentation purposes, the contents of costs is also sent here. This operation would typically occur in the body of getCosts().

## 4.2 The Partition Graph

After profiling, the normalized source files are submitted to the partitioner to assign placements for code and data in the program. First, the partitioner performs an object-sensitive points-to analysis [22] using the Accrue Analysis Framework [7]. This analysis approximates the set of objects that may be referenced by each expression in the program. Using the results of the points-to analysis, an interprocedural def/use analysis links together all assignment statements (defs) with expressions that may observe the value of those assignments (uses). Next, a control dependency analysis [3] links statements that cause branches in the control flow graph (i.e., ifs, loops, and calls) with the statements whose execution depends on them. For instance, each statement in the body of a loop has a control dependency on the loop condition and is therefore linked to the loop condition.

The precision of these analyses can affect the quality of the partitions found by Pyxis and therefore performance. To preserve soundness, the analysis is *conservative*, which means that some dependencies identified by the analysis may not be necessary. Unnecessary dependencies result in inaccuracies in the cost model and superfluous data transfers at run time. For this work we used a "2full+1H" object-sensitive analysis as described in [29].

**Dependencies.** Using these analyses, the partitioner builds the partition graph, which represents information about the program's dependencies. The partition graph contains nodes for each statement in the program and edges for the dependencies between them.

In the partition graph, each statement and field in the program is represented by a node in the graph. Dependencies between statements and fields are represented by edges, and edges have weights that model the cost of satisfying those dependencies. Edges represent different kinds of dependencies between statements:

• A *control edge* indicates a control dependency between two nodes in which the computation at the source node influences whether the statement at the destination node is executed.

• A *data edge* indicates a data dependency in which a value assigned in the source statement influences the computation performed at the destination statement. In this case the source statement is a *definition* (or *def*) and the destination is a *use*.

• An *update edge* represents an update to the heap and connects field declarations to statements that update them.

Part of the partition graph for our running example is shown in Fig. 4. Each node in the graph is labeled with the corresponding line number from Fig. 2. Note, for example, that while lines 20–22 appear sequentially in the program text, the partition graph shows that these lines can be safely executed in any order, as long as they follow line 19. The partitioner adds additional edges (not shown) for output dependencies (write-after-write) and anti-dependencies (read-before-write), but these edges are currently only used during code generation and do not affect the choice of node placements. For each JDBC call that interacts with the database, we also insert control edges to nodes representing "database code."

**Edge weights.** The tradeoff between network overhead and server load is represented by weights on nodes and edges. Each statement node assigned to the database results in additional estimated server load in proportion to the execution count of that statement. Likewise, each dependency that connects two statements (or a statement and a field) on separate partitions incurs estimated network latency proportional to the number of times the dependency is satisfied. Let $\mathrm{cnt}(s)$ be the number of times statement $s$ was executed in the profile. Given a control or data edge from statement $src$ to statement $dst$, we approximate the number of times the edge $e$ was satisfied as $\mathrm{cnt}(e) = \min(\mathrm{cnt}(src), \mathrm{cnt}(dst))$.

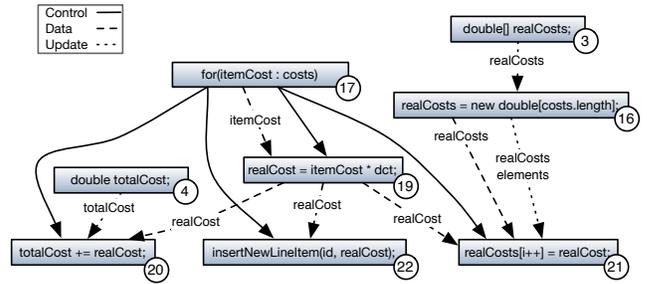

**Figure 4: A partition graph for part of the code from Fig. 2**

Let $\mathrm{size}(def)$ represent the average size of the data that is assigned by statement $def$. Then, given an average network latency LAT and a bandwidth BW, the partitioner assigns weights to edges $e$ and nodes $s$ as follows:

- Control edge $e$: $\mathrm{LAT} \cdot \mathrm{cnt}(e)$

- Data edge $e$: $\dfrac{\mathrm{size}(src)}{\mathrm{BW}} \cdot \mathrm{cnt}(e)$

- Update edge $e$: $\dfrac{\mathrm{size}(src)}{\mathrm{BW}} \cdot \mathrm{cnt}(dst)$

- Statement node $s$: $\mathrm{cnt}(s)$

- Field node: 0

Note that some of these weights, such as those on control and data edges, represent times. The partitioner's objective is to minimize the sum of the weights of edges cut by a given partitioning. The weight on statement nodes is used separately to enforce the constraint that the total CPU load on the server does not exceed a given maximum value.

The formula for data edges charges for bandwidth but not latency. For all but extremely large objects, the weights of data edges are therefore much smaller than the weights of control edges. By setting weights this way, we leverage an important aspect of the Pyxis runtime: satisfying a data dependency does not necessarily require separate communication with the remote host. As described in Sec. 3.2, PyxIL programs maintain consistency of the heap by batching updates and sending them on control transfers. Control dependencies between statements on different partitions inherently require communication in order to transfer control. However, since updates to data can be piggy-backed on a control transfer, the marginal cost to satisfy a data dependency is proportional to the size of the data. For most networks, bandwidth delay is much smaller than propagation delay, so reducing the number of messages a partition requires will reduce the average latency even though the size of messages increases. Furthermore, by encoding this property of data dependencies as weights in the partition graph, we influence the choice of partition made by the solver; cutting control edges is usually more expensive than cutting data edges.

Our simple cost model does not always accurately estimate the cost of control transfers. For example, a series of statements in a block may have many control dependencies to code outside the block. Cutting all these edges could be achieved with as few as one control transfer at runtime, but in the cost model, each cut edge

$$\text{Minimize:} \sum_{e_i \in \text{Edges}} e_i \cdot w_i$$

$$\text{Subject to:} \quad \begin{aligned} n_j - n_k - e_i &\leq 0 \\ n_k - n_j - e_i &\leq 0 \\ &\vdots \end{aligned}$$

$$\sum_{n_i \in \text{Nodes}} n_i \cdot w_i \leq \text{Budget}$$

**Figure 5: Partitioning problem**

contributes its weight, leading to overestimation of the total partitioning cost. Also, fluctuations in network latency and CPU utilization could also lead to inaccurate average estimates and thus result in suboptimal partitions. We leave more accurate cost estimates to future work.

### 4.3 Optimization Using Integer Programming

The weighted graph is then used to construct a Binary Integer Programming problem [31]. For each node we create a binary variable $n \in$ Nodes that has value 0 if it is partitioned to the application server and 1 if it is partitioned to the database. For each edge we create a variable $e \in$ Edges that is 0 if it connects nodes assigned to the same partition and 1 if it is *cut*; that is, the edge connects nodes on different partitions. This problem formulation seeks to minimize network latency subject to a specified budget of instructions that may be executed on the server. In general, the problem has the form shown in Figure 5. The objective function is the summation of edge variables $e_i$ multiplied by their respective weights, $w_i$. For each edge we generate two constraints that force the edge variable $e_i$ to equal 1 if the edge is cut. Note that for both of these constraints to hold, if $n_j \neq n_k$ then $e_i = 1$, and if $n_j = n_k$ then $e_i = 0$. The final constraint in Figure 5 ensures that the summation of node variables $n_i$, multiplied by their respective weights $w_i$, is at most the "budget" given to the partitioner. This constraint limits the load assigned to the database server.

In addition to the constraints shown in Figure 5, we add placement constraints that pin certain nodes to the server or the client. For instance, the "database code" node used to model the JDBC driver's interaction with the database must always be assigned to the database, and similarly assign code that prints on the user's console to the application server.

The placement constraints for JDBC API calls are more interesting. Since the JDBC driver maintains unserializable native state regarding the connection, prepared statements, and result sets used by the program, all API calls must occur on the same partition. While these calls could also be pinned to the database, this could result in low-quality partitions if the partitioner has very little budget. Consider an extreme case where the partitioner is given a budget of 0. Ideally, it should create a partition equivalent to the original program where all statements are executed on the application server. Fortunately, this behavior is easily encoded in our model by assigning the same node variable to all statements that contain a JDBC call and subsequently solving for the values of the node variables. This encoding forces the resulting partition to assign all such calls to the same partition.

After instantiating the partitioning problem, we invoke the solver. If the solver returns with a solution, we apply it to the partition graph by assigning all nodes a location and marking all edges that are cut. Our implementation currently supports lpsolve [20] and Gurobi Optimizer[16].

Finally, the partitioner generates a PyxIL program from the partition graph. For each field and statement, the code generator emits a placement annotation `:APP:` or `:DB:` according to the solution returned by the solver. For each dependency edge between remote statements, the code generator places a heap synchronization operation after the source statement to ensure the remote heap is up to date when the destination statement is executed. Synchronization operations are always placed after statements that update remotely partitioned fields or arrays.

### 4.4 Statement Reordering

A partition graph may generate several valid PyxIL programs that have the same cost under the cost model since the execution order of some statements in the graph is ambiguous. To eliminate unnecessary control transfers in the generated PyxIL program, the code generator performs a reordering optimization to create larger contiguous blocks with the same placement, reducing control transfers. Because it captures all dependencies, the partition graph is particularly well suited to this transformation. In fact, PDGs have been applied to similar problems such as vectorizing program statements [14]. The reordering algorithm is simple. Recall that in addition to control, data, and update edges, the partitioner includes additional (unweighted) edges for output dependencies (for ordering writes) and anti-dependencies (for ordering reads before writes) in the partition graph. We can usefully reorder the statements of each block without changing the semantics of the program [14] by topologically sorting the partition graph while ignoring back-edges and interprocedural edges[4].

The topological sort is implemented as a breadth-first traversal over the partition graph. Whereas a typical breadth-first traversal would use a single FIFO queue to keep track of nodes not yet visited, the reordering algorithm uses two queues, one for `DB` statements and one for `APP` statements. Nodes are dequeued from one queue until it is exhausted, generating a sequence of statements that are all located in one partition. Then the algorithm switches to the other queue and starts generating statements for the other partition. This process alternates until all nodes have been visited.

### 4.5 Insertion of Synchronization Statements

The code generator is also responsible for placing heap synchronization statements to ensure consistency of the Pyxis distributed heap. Whenever a node has outgoing data edges, the code generator emits a `sendAPP` or `sendDB` depending on where the updated data is partitioned. At the next control transfer, the data for all such objects so recorded is sent to update of the remote heap.

Heap synchronization is conservative. A data edge represents a definition that may reach a field or array access. Eagerly synchronizing each data edge ensures that all heap locations will be up to date when they are accessed. However, imprecision in the reaching definitions analysis is unavoidable, since predicting future accesses is undecidable. Therefore, eager synchronization is sometimes wasteful and may result in unnecessary latency from transferring updates that are never used.

The Pyxis runtime system also supports lazy synchronization in which objects are fetched from the remote heap at the point of use. If a use of an object performs an explicit fetch, data edges to that use can be ignored when generating data synchronization. Lazy synchronization makes sense for large objects and for uses in infrequently executed code. In the current Pyxis implementation, lazy synchronization is not used in normal circumstances. We leave a hybrid eager/lazy synchronization scheme to future work.

---

[4]Side-effects and data dependencies due to calls are summarized at the call site.

```
1  public class Order {
2    ObjectId oid;
3    class Order_app { int id; ObjectId realCostsId; }
4    class Order_db { double totalCost; }
5    ...
6  }
```
**Figure 6: Partitioning fields into APP and DB objects**

## 5. PYXIL COMPILER

The PyxIL compiler translates PyxIL source code into two partitioned Java programs that together implement the semantics of the original application when executed on the Pyxis runtime. To illustrate this process, we give an abridged version of the compiled `Order` class from the running example. Fig. 6 shows how `Order` objects are split into two objects of classes `Order_app` and `Order_db`, containing the fields assigned to APP and DB respectively in Fig. 3. Both classes contain a field `oid` (line 2) which is used to identify the object in the Pyxis-managed heaps (denoted by `DBHeap` and `APPHeap`).

### 5.1 Representing Execution Blocks

In order to arbitrarily assign program statements to either the application or the database server, the runtime needs to have complete control over program control flow as it transfers between the servers. Pyxis accomplishes this by compiling each PyxIL method into a set of *execution blocks*, each of which corresponds to a straight-line PyxIL code fragment. For example, Fig. 7 shows the code generated for the method `computeTotalCost` in the class. This code includes execution blocks for both the APP and DB partitions. Note that local variables in the PyxIL source are translated into indices of an array `stack` that is explicitly maintained in the Java code and is used to model the stack frame that the method is currently operating on.

The key to managing control flow is that each execution block ends by returning the identifier of the next execution block. This style of code generation is similar to that used by the SML/NJ compiler [30] to implement continuations; indeed, code generated by the PyxIL compiler is essentially in continuation-passing style [13].

For instance, in Fig. 7, execution of `computeTotalCost` starts with block `computeTotalCost0` at line 1. After creating a new stack frame and passing in the object ID of the receiver, the block asks the runtime to execute `getCosts0` with `computeTotalCost1` recorded as the return address. The runtime then executes the execution blocks associated with `getCosts()`. When `getCosts()` returns, it jumps to `computeTotalCost1` to continue the method, where the result of the call is popped into `stack[2]`.

Next, control is transferred to the database server on line 14, because block `computeTotalCost1` returns the identifier of a block that is assigned to the database server (`computeTotalCost2`). This implements the transition in Fig. 3 from (APP) line 21 to (DB) line 22. Execution then continues with `computeTotalCost3`, which implements the evaluation of the loop condition in Fig. 3.

This example shows how the use of execution blocks gives the Pyxis partitioner complete freedom to place each piece of code and data to the servers. Furthermore, because the runtime regains control after every execution block, it has the ability to perform other tasks between execution blocks or while waiting for the remote server to finish its part of the computation such as garbage collection on local heap objects.

### 5.2 Interoperability with Existing Modules

Pyxis does not require that the whole application be partitioned. This is useful, for instance, if the code to be partitioned is a library used by other, non-partitioned code. Another benefit is that code without an explicit main method can be partitioned, such as a servlet whose methods are invoked by the application server.

To use this feature, the developer indicates the *entry points* within the source code to be partitioned, i.e., methods that the developer exposes to invocations from non-partitioned code. The PyxIL compiler automatically generates a wrapper for each entry point, such as the one shown in Fig. 8, that does the necessary stack setup and teardown to interface with the Pyxis runtime.

## 6. PYXIS RUNTIME SYSTEM

The Pyxis runtime executes the compiled PyxIL program. The runtime is a Java program that runs in unmodified JVMs on each server. In this section we describe its operation and how control transfer and heap synchronization is implemented.

### 6.1 General Operations

The runtime maintains the program stack and distributed heap. Each execution block discussed in Sec. 5.1 is implemented as a Java class with a `call` method that implements the program logic of the given block. When execution starts from any of the entry points, the runtime invokes the `call` method on the block that was passed in from the entry point wrapper. Each `call` method returns the next block for the runtime to execute next, and this process continues on the local runtime until it encounters an execution block that is assigned to the remote runtime. When that happens, the runtime sends a *control transfer message* to the remote runtime and waits until the remote runtime returns control, informing it of the next execution block to run on its side.

Pyxis does not currently support thread instantiation or shared-memory multithreading, but a multithreaded application can first instantiate threads outside of Pyxis and then have Pyxis manage the code to be executed by each of the threads. Additionally, the current implementation does not support exception propagation across servers, but extending support in the future should not require substantial engineering effort.

### 6.2 Program State Synchronization

When a control transfer happens, the local runtime needs to communicate with the remote runtime about any changes to the program state (i.e., changes to the stack or program heap). Stack changes are always sent along with the control transfer message and as such are not explicitly encoded in the PyxIL code. However, requests to synchronize the heaps are explicitly embedded in the PyxIL code, allowing the partitioner to make intelligent decisions about what modified objects need to be sent and when. As mentioned in Sec. 4.5, Pyxis includes two heap synchronization routines: `sendDB` and `sendAPP`, depending on which portion of the heap is to be sent. In the implementation, the heap objects to be sent are simply piggy-backed onto the control transfer messages (just like stack updates) to avoid initiating more round trips. We measure the overhead of heap synchronization and discuss the results in Sec. 7.3.

### 6.3 Selecting a Partitioning Dynamically

The runtime also supports dynamically choosing between partitionings with different CPU budgets based on the current load on the database server. It uses a feedback-based approach in which the Pyxis runtime on the database server periodically polls the CPU utilization on the server and communicates that information to the application server's runtime.

At each time $t$ when a load message arrives with server load $S_t$, the application server computes a weighted moving average (EWMA) of the load, $L_t = \alpha L_{t-1} + (1 - \alpha) S_t$. Depending

1477

```
    stack locations for computeTotalCost:
    stack[0] = oid
    stack[1] = dct
    stack[2] = object ID for costs
    stack[3] = costs.length
    stack[4] = i
    stack[5] = loop index
    stack[6] = realCost
```

```
1   computeTotalCost0:
2     pushStackFrame(stack[0]);
3     setReturnPC(computeTotalCost1);
4     return getCosts0; // call this.getCosts()
5
6   computeTotalCost1:
7     stack[2] = popStack();
8     stack[3] = APPHeap[stack[2]].length;
9     oid = stack[0];
10    APPHeap[oid].realCosts = nativeObj(new dbl[stack[3]]);
11    sendAPP(oid);
12    sendNative(APPHeap[oid].realCosts, stack[2]);
13    stack[4] = 0;
14    return computeTotalCost2; // control transfer to DB
```

```
15  computeTotalCost2:
16    stack[5] = 0; // i = 0
17    return computeTotalCost3; // start the loop
18
19  computeTotalCost3:
20    if (stack[5] < stack[3]) // loop index < costs.length
21        return computeTotalCost4; // loop body
22    else return computeTotalCost5; // loop exit
23
24  computeTotalCost4:
25    itemCost = DBHeap[stack[2]][stack[5]];
26    stack[6] = itemCost * stack[1];
27    oid = stack[0];
28    DBHeap[oid].totalCost += stack[6];
29    sendDB(oid);
30    APPHeap[oid].realCosts[stack[4]++] = stack[6];
31    sendNative(APPHeap[oid].realCosts);
32    ++stack[5];
33    pushStackFrame(APPHeap[oid].id, stack[6]);
34    setReturnPC(computeTotalCost3);
35    return insertNewLineItem0; // call insertNewLineItem
36
37  computeTotalCost5:
38    return returnPC;
```

**Figure 7: Running example: APP code (left) and DB code (right). For simplicity, execution blocks are presented as a sequence of statements preceded by a label.**

```
public class Order {
  ...
  public void computeTotalCost(double dct) {
    pushStackFrame(oid, dct);
    execute(computeTotalCost0);
    popStackFrame();
    return; // no return value
  }
}
```

**Figure 8: Wrapper for interfacing with regular Java**

on the value of $L_t$, the application server's runtime dynamically chooses which partition to execute at each of the entry points. For instance, if $L_t$ is high (i.e., the database server is currently loaded), then the application server's runtime will choose a partitioning that was generated using a low CPU budget until the next load message arrives. Otherwise the runtime uses a partitioning that was generated with a higher CPU-budget since $L_t$ indicates that CPU resources are available on the database server. The use of EWMA here prevents oscillations from one deployment mode to another. In our experiments with TPC-C (Sec. 7.1.3), we used two different partitions and set the threshold between them to be 40% (i.e., if $L_t > 40$ then the runtime uses a lower CPU-budget partition). Load messages were sent every 10 seconds with $\alpha$ set to 0.2. The values were determined after repeat experimentation.

This simple, dynamic approach works when separate client requests are handled completely independently at the application server, since the two instances of the program do not share any state. This scenario is likely in many server settings; e.g., in a typical web server, each client is completely independent of other clients. Generalizing this approach so that requests sharing state can utilize this adaptation feature is future work.

## 7. EXPERIMENTS

In this section we report experimental results. The goals of the experiments are: to evaluate Pyxis's ability to generate partitions of an application under different server loads and input workloads; and to measure the performance of those partitionings as well as the overhead of performing control transfers between runtimes. Pyxis is implemented in Java using the Polyglot compiler framework [24] with Gurobi and lpsolve as the linear program solvers. The experiments used mysql 5.5.7 as the DBMS with buffer pool size set to 1GB, hosted on a machine with 16 2.4GHz cores and 24GB of physical RAM. We used Apache Tomcat 6.0.35 as the web server for TPC-W, hosted on a machine with eight 2.6GHz cores and 33GB of physical RAM. The disks on both servers are standard serial ATA disks. The two servers are physically located in the same data center and have a ping round trip time of 2ms. All reported performance results are the average of three experimental runs.

For TPC-C and TPC-W experiments below, we implemented three different versions of each benchmark and measured their performance as follows:

- **JDBC**: This is a standard implementation of each benchmark where program logic resides completely on the application server. The program running on the application server connects to the remote DBMS using JDBC and makes requests to fetch or write back data to the DBMS using the obtained connection. A round trip is incurred for each database operation between the two servers.

- **Manual**: This is an implementation of the benchmarks where all the program logic is manually split into two halves: the "database program," which resides on the JVM running on the database server, and the "application program," which resides on the application server. The application program is simply a wrapper that issues RPC calls via Java RMI to the database program for each type of transaction, passing along with it the arguments to each transaction type. The database program executes the actual program logic associated with each type of transaction and opens local connections to the DBMS for the database operations. The final results are returned to the application program. This is the exact opposite from the JDBC implementation with all program logic residing on the database server. Here, each transaction only incurs one round trip. For TPC-C we also implemented a version that implements each transaction as a MySQL user-defined function rather than issuing JDBC calls from a Java program on the database server. We found that this did not significantly impact the performance results.

- **Pyxis**: To obtain instruction counts, we first profiled the JDBC implementation under different target throughput rates for a fixed period of time. We then asked Pyxis to generate different partitions with different CPU budgets. We deployed the two partitions on the application and database servers using Pyxis and measured their performance.

## 7.1 TPC-C Experiments

In the first set of experiments we implemented the TPC-C workload in Java. Our implementation is similar to an "official" TPC-C implementation but does not include client think time. The database contains data from 20 warehouses (initial size of the database is 23GB), and for the experiments we instantiated 20 clients issuing new order transactions simultaneously from the application server to the database server with 10% transactions rolled back. We varied the rate at which the clients issued the transactions and then measured the resulting system's throughput and average latency of the transactions for 10 minutes.

### 7.1.1 Full CPU Setting

In the first experiment we allowed the DBMS and JVM on the database server to use all 16 cores on the machine and gave Pyxis a large CPU budget. Fig. 9(a) shows the throughput versus latency. Fig. 9(b) and (c) show the CPU and network utilization under different throughputs.

The results illustrate several points. First, the Manual implementation was able to scale better than JDBC both in terms of achieving lower latencies, and being able to process more transactions within the measurement period (i.e., achieve a high overall throughput). The higher throughput in the Manual implementation is expected since each transaction takes less time to process due to fewer round trips and incurs less lock contention in the DBMS due to locks being held for less time.

For Pyxis, the resulting partitions for all target throughputs were very similar to the Manual implementation. Using the provided CPU budget, Pyxis assigned most of the program logic to be executed on the database server. This is the desired result; since there are CPU resources available on the database server, it is advantageous to push as much computation to it as possible to achieve maximum reduction in the number of round trips between the servers. The difference in performance between the Pyxis and Manual implementations is negligible (within the margin of error of the experiments due to variation in the TPC-C benchmark's randomly generated transactions).

There are some differences in the operations of the Manual and Pyxis implementations, however. For instance, in the Manual implementation, only the method arguments and return values are communicated between the two servers whereas the Pyxis implementation also needs to transmit changes to the program stack and heap. This can be seen from the network utilization measures in Fig. 9(c), which show that the Pyxis implementation transmits more data compared to the Manual implementation due to synchronization of the program stack between the runtimes. We experimented with various ways to reduce the number of bytes sent, such as with compression and custom serialization, but found that they used more CPU resources and increased latency. However, Pyxis sends data only during control transfers, which are fewer than database operations. Consequently, Pyxis sends less data than the JDBC implementation.

### 7.1.2 Limited CPU Setting

Next we repeated the same experiment, but this time limited the DBMS and JVM (applicable only to Manual and Pyxis implementations) on the database server to use a maximum of three CPUs and gave Pyxis a small CPU budget. This experiment was designed to emulate programs running on a highly contended database server or a database server serving load on behalf of multiple applications or tenants. The resulting latencies are shown in Fig. 10(a) with CPU and network utilization shown in Fig. 10(b) and (c). The Manual implementation has lower latencies than Pyxis and JDBC when the throughput rate is low, but for higher throughput values the JDBC and Pyxis implementations outperforms Manual. With limited CPUs, the Manual implementation uses up all available CPUs when the target throughput is sufficiently high. In contrast, all partitions produced by Pyxis for different target throughput values resemble the JDBC implementation in which most of the program logic is assigned to the application server. This configuration enables the Pyxis implementation to sustain higher target throughputs and deliver lower latency when the database server experiences high load. The resulting network and CPU utilization are similar to those of JDBC as well.

### 7.1.3 Dynamically Switching Partitions

In the final experiment we enabled the dynamic partitioning feature in the runtime as described in Sec. 6.3. The two partitionings used in this case were the same as the partitionings used in the previous two experiments (i.e., one that resembles Manual and another that resembles JDBC). In this experiment, however, we fixed the target throughput to be 500 transactions / second for 10 minutes since all implementations were able to sustain that amount of throughput. After three minutes elapsed we loaded up most of the CPUs on the database to simulate the effect of limited CPUs. The average latencies were then measured during each 30-second period, as shown in Fig. 11. For the Pyxis implementation, we also measured the proportion of transactions executed using the JDBC-like partitioning within each 1-minute period. Those are plotted next to each data point.

As expected, the Manual implementation had lower latencies when the server was not loaded. For JDBC the latencies remain constant as it does not use all available CPUs even when the server is loaded. When the server is unloaded, however, it had higher latency compared to the Manual implementation. The Pyxis implementation, on the other hand, was able to take advantage of the two implementations with automatic switching. In the ideal case, Pyxis's latencies should be the minimum of the other two implementations at all times. However, due to the use of EWMA, it took a short period of time for Pyxis to adapt to load changes, although it eventually settled to an all-application (JDBC-like) deployment as shown by the proportion numbers. This experiment illustrates that even if the developer was not able to predict the amount of available CPU resources, Pyxis can generate different partitions under various budgets and automatically choose the best one given the actual resource availability.

## 7.2 TPC-W

In the next set of experiments, we used a TPC-W implementation written in Java. The database contained 10,000 items (about 1GB on disk), and the implementation omitted the thinking time. We drove the load using 20 emulated browsers under the browsing mix configuration and measured the average latencies at different target Web Interactions Per Seconds (WIPS) over a 10-minute period. We repeated the TPC-C experiments by first allowing the JVM and DBMS to use all 16 available cores on the database server followed by limiting to three cores only. Fig. 12 and Fig. 13 show the latency results. The CPU and network utilization are similar to those in the TPC-C experiments and are not shown.

Compared to the TPC-C results, we see a similar trend in latency. However, since the program logic in TPC-W is more complicated than TPC-C, the Pyxis implementation incurs a bit more overhead compared to the Manual implementation.

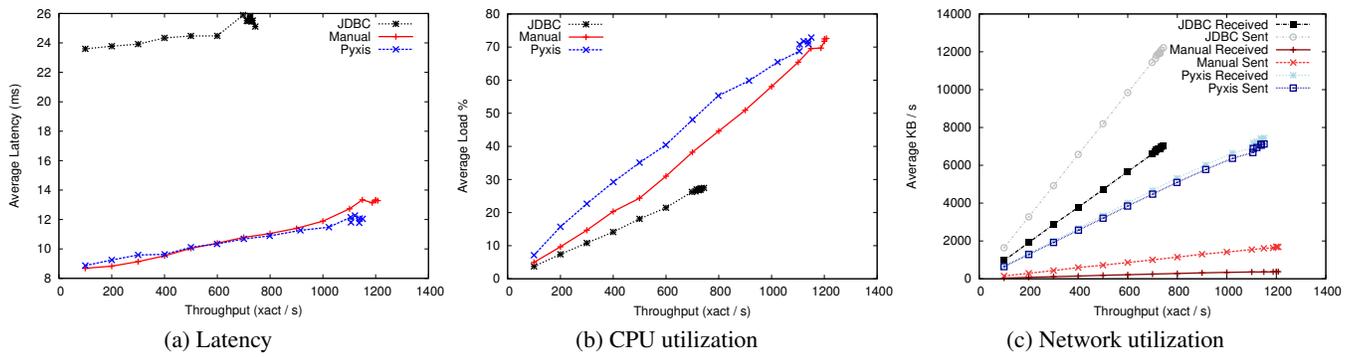

Figure 9: TPC-C experiment results on 16-core database server

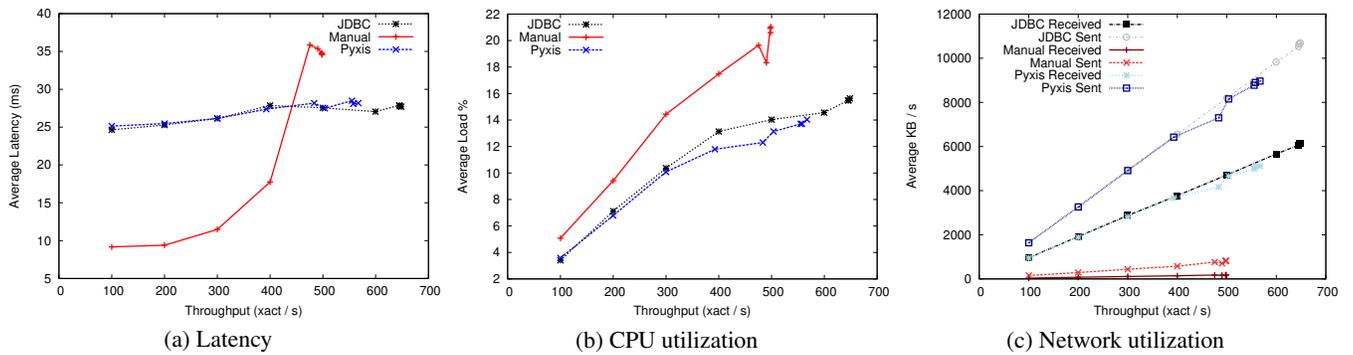

Figure 10: TPC-C experiment results on 3-core database server

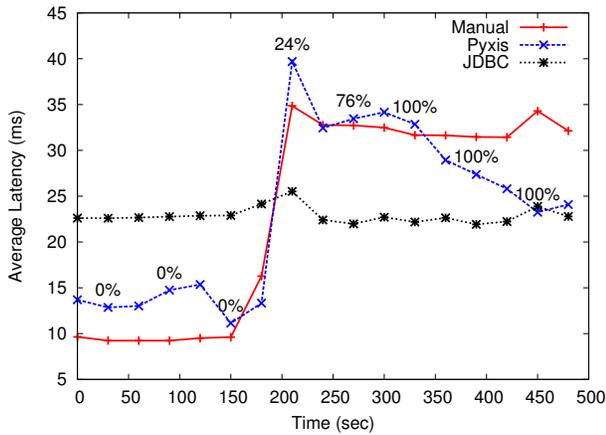

Figure 11: TPC-C performance results with dynamic switching

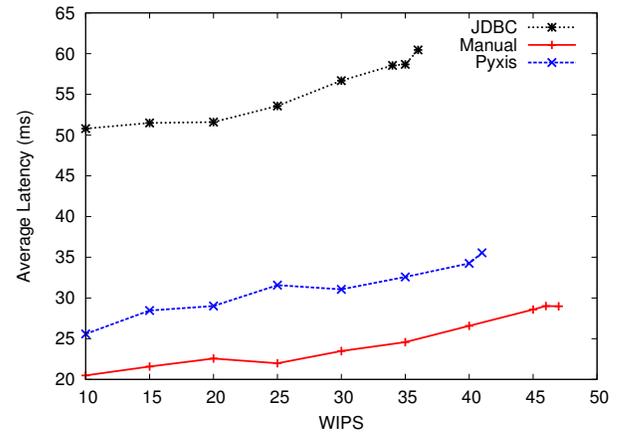

Figure 12: TPC-W latency experiment using 16 cores

One interesting aspect of TPC-W is that unlike TPC-C, some web interactions do not involve database operations at all. For instance, the order inquiry interaction simply prints out a HTML form. Pyxis decides to place the code for those interactions entirely on the application server even when the server budget is high. This choice makes sense since executing such interactions on the application server does not incur any round trips to the database server. Thus the optimal decision, also found by Pyxis, is to leave the code on the application server rather than pushing it to the database server as stored procedures.

## 7.3 Microbenchmark 1

In our third experiment we compared the overhead of Pyxis to native Java code. We expected code generated by Pyxis to be slower because all heap and stack variable manipulations in Pyxis are managed through special Pyxis objects. To quantify this overhead we implemented a linked list and assigned all the fields and code to be on the same server. This configuration enabled a fair comparison to a native Java implementation. The results show that the Pyxis implementation has an average overhead of 6× compared to the Java implementation. Since the experiment did not involve any control transfers, the overhead is entirely due to the time spent running execution blocks and bookkeeping in the Pyxis managed program stack and heap. This benchmark is a worst-case scenario for Pyxis since the application is not distributed at all. We expect that programmers will ask Pyxis to partition fragments of an application that involve database operations and will implement the rest of their program as native Java code. As noted in Sec. 5.2, this separation of distributed from local code is fully supported in our implementation; programmers simply need to mark the classes or functions they want to partition. Additionally, results in the previous sections show that allowing Pyxis to decide on the partitioning for distributed applications offers substantial benefits over native Java (e.g., the JDBC and Manual implementations) even though Pyxis pays some local execution overhead.

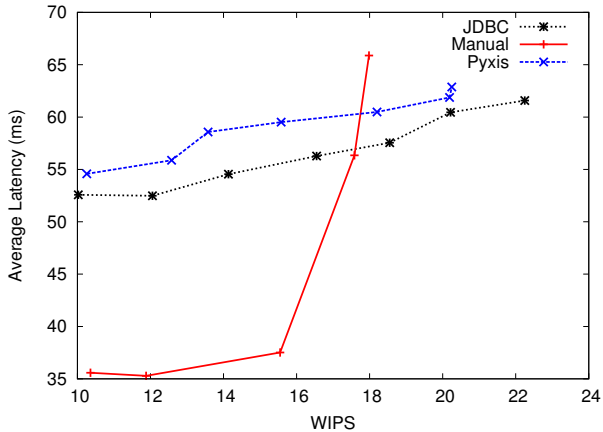

**Figure 13: TPC-W latency experiment using 3 cores**

| CPU Load | APP | APP—DB | DB |
|---|---|---|---|
| No load | 7m37s | 7m | **6m39s** |
| Partial load | 8m55s | **8m** | 8m11s |
| Full load | **16m36s** | 17m30s | 17m44s |

**Figure 14: Microbenchmark 2 results**

The overhead of Pyxis is hurt by the mismatch between the Java Virtual Machine execution model and Pyxis's execution model based on execution blocks. Java is not designed to make this style of execution fast. One concrete example of this problem is that we cannot use the Java stack to store local variables. Targeting a lower-level language should permit lowering the overhead substantially.

## 7.4 Microbenchmark 2

In the final experiment, we compare the quality of the generated partitions under different CPU budgets by using a simple microbenchmark designed to have several interesting partitionings. This benchmark runs three tasks in order: it issues a 100k small select queries, performs a computationally intensive task (compute SHA1 digest 500k times), and issues another 100k select queries. We gave three different CPU budget values (low, medium, high) to represent different server loads and asked Pyxis to produce partitions. Three different partitions were generated: one that assigns all logic to application server when a low budget was given (**APP**); one that assigns the query portions of the program to the database server and the computationally intensive task to the application when a middle range budget was given (**APP—DB**); and finally one that assigns everything to the database server (**DB**). We measured the time taken to run the program under different real server loads, and the results are shown in Fig. 14.

The results show that Pyxis was able to generate a partition that fits different CPU loads in terms of completion time (highlighted in Fig. 14). While it might be possible for developers to avoid using the Pyxis partitioner and manually create the two extreme partitions (**APP** and **DB**), this experiment shows that doing so would miss the "middle" partitions (such as **APP—DB**) and lead to suboptimal performance. Using Pyxis, the developer only needs to write the application once, and Pyxis will automatically produce partitions that are optimized for different server loads.

## 8. RELATED WORK

Several prior systems have explored automatically partitioning applications across distributed systems. However, Pyxis is the first system that partitions general database applications between an application server and a database server. Pyxis also contributes new techniques for reducing data transfers between host nodes.

**Program partitioning.** Program partitioning has been an active research topic for the past decade. Most of these approaches require programs to be decomposed into coarse-grained modules or functions that simplify and make a program's dependencies explicit. Imposing this structure reduces the complexity of the partitioning problem and has been usefully applied to several tasks: constructing pages in web applications with Hilda [33], processing data streams in sensor networks with Wishbone [23], and optimizing COM applications with Coign [18].

Using automatic partitioning to offload computation from handheld devices has featured in many recent projects. Odessa [25] dynamically partitions computation intensive mobile applications but requires programs to be structured as pipelined processing components. Other projects such as MAUI [12] and CloneCloud [9] support more general programs, but only partition at method boundaries. Chroma [4] and Spectra [4] also partition at method boundaries but improve partition selection using code annotations called *tactics* to describe possible partitionings.

Prior work on secure program partitioning such as Swift [8] and Jif/split [34, 35] focuses more on security than performance. For instance, Swift minimizes control transfers but does not try to optimize data transfers.

Pyxis supports a more general programming model compared to these approaches. Our implementation targets Java programs and the JDBC API, but our technique is applicable to other general purpose programming languages. Pyxis does not require special structuring of the program nor does it require developer annotations. Furthermore, by considering program dependencies at a fine-grained level, Pyxis can automatically create code and data partitions that would require program refactoring or design changes to affect in other systems.

**Data partitioning and query rewriting.** Besides program partitioning, another method to speed up database applications is to rewrite or batch queries embedded in the application. One technique for automatically restructuring database workloads is to issue queries asynchronously, using program analysis to determine when concurrent queries do not depend on each other [6], or using data dependency information to batch queries together [15]. Automatic partitioning has advantages over this approach: it works even in cases where successive queries do depend on each other, and it can reduce data transfers between the hosts.

Data partitioning has been widely studied in the database research community [2, 27], but the techniques are very different. These systems focus on distributed query workloads and only partition data among different database servers, whereas Pyxis partitions both data and program logic between the database server and its client.

**Data persistence languages.** Several projects explore integrated query languages [32, 11, 21, 10, 19] that improve expressiveness and performance of database applications. Pyxis partitions database applications that use the standard JDBC API. However, partitioning programs that employ integrated query languages would ease analysis of the queries themselves and could result in additional optimizations. Raising the level of abstraction for database interaction would also allow Pyxis to overcome difficulties like constraining JDBC API calls to be on the same partition, and can potentially improve the quality of the generated partitions.

**Custom language runtimes.** Sprint [26] speculatively executes branches of a program to predict future accesses of remote data and reduce overall latency by prefetching. The Sprint system is targeted at read-only workloads and performs no static analysis of the program to predict data dependencies. Pyxis targets a more general set of applications and does not require speculative execution to conservatively send data between hosts before the data is required.

Executable program slicing [5] extracts the program fragment that a particular program point depends on using the a system dependence graph. Like program slicing, Pyxis uses a control and data dependencies to extract semantics-preserving fragments of programs. The code blocks generated by Pyxis, however, are finer-grained fragments of the original program than a program slice, and dependencies are satisfied by automatically inserting explicit synchronization operations.

## 9. CONCLUSIONS

In this paper we presented Pyxis, a system for partitioning database applications across an application and database server. The goal of this partitioning is to eliminate the many small, latency-inducing round trips between the application and database servers by pushing blocks of code into the database as stored procedures. Benefits of our approach include that users are not required to manually convert their code into the stored procedure representation, and that our system can dynamically identify the best decomposition of the application based on variations in server load. Our key technical contributions include a cost-based profiling and analysis framework, as well as a compilation strategy for converting the output of our analysis into valid executables. Our experiments show that Pyxis is able to provide performance benefits that are as good as a manual conversion to stored procedures, providing up to $3\times$ reduction in latency and a $1.7\times$ improvement in throughput versus a conventional JDBC-based implementation of TPC-C.